\documentclass[preprintnumbers, twocolumn, preprintnumbers, PRD, superscriptaddress, nofootinbib]{revtex4-2}
\usepackage{hyperref}
\hypersetup{
  colorlinks=true,        
  linkcolor=blue,         
  citecolor=cyan,         
}
\usepackage{cancel}
\usepackage{amsfonts}
\usepackage{amsmath}
\usepackage{amssymb,epsf}
\usepackage{latexsym}
\usepackage{graphicx,epsfig}
\usepackage{amssymb}
\usepackage{float}
\usepackage{subfigure}
\usepackage{epstopdf}
\usepackage{dcolumn}
\usepackage{psfrag}
\usepackage{wrapfig}
\usepackage{makeidx}
\usepackage{epsf}
\usepackage{color}
\usepackage{multirow}
\usepackage{tabularx}
\usepackage{array}
\usepackage{physics}
\numberwithin{equation}{section}
\def\be {\begin{equation}}
\def\ee {\end{equation}}
\def\bea {\begin{eqnarray}}
\def\eea {\end{eqnarray}}
\def\bc {\begin{center}}
\def\ec {\end{center}}
\def\bfg {\begin{figure}}
\def\efg {\end{figure}}
\def\bi {\begin{itemize}}
\def\ei {\end{itemize}}

\def\beq{\begin{equation}}
\def\eeq{\end{equation}}
\def\br{\begin{eqnarray}}
\def\er{\end{eqnarray}}
\newcommand{\eel}[1] {\label{#1}\end{equation}}

\begin{document}
\title{\Large Exploring a novel Einstein--Rosen BTZ wormhole}
\author{Ankit Anand}
\email{ankitanandp94@gmail.com} 
\affiliation{Physics Division, School of Basic and Applied Sciences, Galgotias University, Greater Noida 203201, India}
\author{Kimet Jusufi}
\email{kimet.jusufi@unite.edu.mk} 
\affiliation{Physics Department,
University of Tetova, Ilinden Street nn, 1200, Tetovo, North
Macedonia}
\author{Mendrit Latifi}
\email{ljatifi@thphys.uni-heidelberg.de} 
\affiliation{Institute for Theoretical Physics, Heidelberg University,
Philosphenweg.19, Heidelberg, 69117, Germany}

\begin{abstract}
We introduce a novel Einstein-Rosen BTZ wormhole metric as a solution to the Einstein field equations with a negative cosmological constant and explore in detail its various phenomenological aspects. We show that the wormhole metric is characterized by a horizon at the throat, resembling a black hole horizon. This implies that our wormhole metric describes a one-way traversable wormhole at the throat, with Hawking radiation observed by an observer located at some distance from the wormhole. It is also found the same Hawking temperature using the BTZ-like coordinates and Kruskal-like coordinates. This temperature is invariant not only on the type of coordinates but also the nature of the spin of quantum fields.  Importantly, we find that at the wormhole throat, the spacetime is not a pure vacuum solution, but rather contains an exotic string matter source with negative tension, which may stabilize the wormhole geometry. To this end, we found that the size of the wormhole throat is proportional to the number of quantum bits suggesting a possible implications on ER=EPR. Further we studied the particle dynamics and, finally, we tested the ANEC with a test scalar and vector fields. For the double null-component computed in BTZ coordinates, we found an apparent divergence at the wormhole throat, which is then shown to be regularized by means of Kruskal-like coordinates. The ANEC for such a scalar/vector field is violated at the wormhole throat. 
\end{abstract}
\maketitle

\section{Introduction}

Wormholes represents one of the most exciting General Relativity (GR) solutions. The wormhole is a spacetime structure that can connect two separate worlds or distant areas within the same universe; it is distinguished by a small surface known as the throat.  In 1935, A. Einstein and N. Rosen proposed the Einstein-Rosen (ER) bridge hypothesis of wormholes \cite{Einstein:1935tc}. It has been shown that this solution belongs to the Kruskal extension of the Schwarzschild metric, and the ER bridge cannot be traversed \cite{Fuller:1962zza, Kruskal:1959vx}. Afterward, the field remained silent for over two decades. Interest was reignited in 1957 when J.A. Wheeler and C.W. Misner coined the term `wormhole' opening a new door for explorations and interest in the area of wormhole physics \cite{Misner:1957mt}. It was also shown that violating the null energy criterion may be essential to create traversable wormholes \cite{Visser:1989kh}. On the other hand, the discovery of traversable wormholes dates back to \cite{Ellis:1973yv, Ellis:1979bh, Bronnikov:1973fh, Kodama:1978dw}, with additional insights offered by Morris and Thorne in 1988 \cite{Morris:1988cz}. Since then, different wormhole geometries have been extensively studied, with significant focus not only on the theoretical aspects but also on their possible relation to astrophysical phenomena such as gravitational lensing, shadows, gravity waves, and other aspects when confronted to the astrophysical data. Such investigations are of particular interest and can shed light on these objects and help us detect or distinguish them from black holes \cite{Simpson:2018tsi,Jusufi:2018waj,Jusufi:2020yus,Jusufi:2021lei,Jusufi:2017mav,Bambhaniya:2021ugr,Ohgami:2015nra,Shaikh:2018kfv,Gyulchev:2018fmd,Dai:2019mse,Bambi:2013nla,Simonetti:2020ivl,Blazquez-Salcedo:2020nsa,Khoo:2024yeh,Huang:2023yqd,Azad:2023iju,DeFalco:1,DeFalco:2,DeFalco:3}.

The AdS/CFT correspondence \cite{Maldacena:1997re} exemplifies holographic duality by stating that gravity theory in an anti-de Sitter (AdS) spacetime is comparable to $\mathcal{N} = 4$ super Yang Mill theory on the AdS boundary. Quantum mechanics uses correspondence to geometrize quantities, similar to how general relativity does for conventional physics. The AdS/CFT theory aims to resolve paradoxes in our understanding of quantum gravity, including the information problem in black hole formation and evaporation. General relativity provides insights into gravity force and answers difficult questions in Newtonian gravity. We consider a probe string in AdS space whose two ends have been attached to the boundary. Another intriguing geometrization of quantum entanglement is ER=EPR, based on comparing ER bridges, also known as wormholes, and Einstein-Podolsky-Rosen pairs, or EPR pairs\cite{Maldacena:2013xja}. This probe string is analogous to the quark and anti-quark EPR pair traveling across $\mathcal{N} = 4$ super Yang-Mills fields in the AdS/CFT content. The induced metric on the string worldsheet is identical to that of a two-sided AdS black hole (wormhole) due to uniformly accelerating ends along the boundary, as discussed in \cite{Yeh:2023avs}. This scenario's entanglement of the quarks in the spirit of ER=EPR and the worldsheet wormhole are closely related, as was initially noted in \cite{Jensen:2013ora, Sonner:2013mba}.

Recently, Gao, Jafferis, and Wall \cite{Gao:2016bin} showed a traversable wormhole from a BTZ black hole by adding a time-dependent connection between its asymptotic boundaries. They used the point-splitting approach to construct the one-loop stress-energy tensor. The wormhole can be traversable by correctly selecting the coupling's sign, resulting in a negative vacuum expectation value for the double null component of the stress-energy tensor. A four-dimensional traversable wormhole was studied in \cite{Maldacena:2018gjk} by joining two charged extremal black holes using massless fermions, relying only on local fermion dynamics. A similar approach in \cite{Fu:2018oaq, Marolf:2019ojx, Anand:2020wlk} used a quantum field in $AdS_3$ and $AdS_3 \cross S^1$ with discrete symmetries. They showed that the quantum fields backreact on the geometry and make the wormhole traversable.

It is commonly believed that the Einstein-Rosen bridge is a vacuum solution. However, more detailed mathematical formulations, as pointed out in \cite{Guendelman:2009er, Guendelman:2016bwj, Guendelman:2015wsv}, indicate that this solution requires the presence of a special type of "exotic" matter located at its throat. Specifically, it has been argued that the Einstein-Rosen metric does not satisfy the vacuum Einstein equations at the throat. This, in turn, implies the presence of an ill-defined, non-vanishing "matter" stress-energy tensor term in the Einstein field equations, which was overlooked in the original 1935 paper by Einstein and Rosen.

In this paper, we aim to explore a novel metric form the Einstein-Rosen 'bridge' in $2+1$ dimensional gravity, namely in the context of BTZ black holes which is a solution of the Einstein field equations with a negative cosmological constant. Toward this goal we will introduce a new coordinate transformation to obtain an Einstein-Rosen metric in BTZ -like coordinates and, importantly, in terms of non-singular Kruskal-like coordinates.  We aim to understand more about it's phenomenological aspects, in particular we will demonstrate that such a wormhole requires the presence of a special type of 'exotic' matter located at its throat. Another interesting aspects will be the thermodynamics aspects, namely the wormhole throat will be shown to play the role of the black hole horizon. This means that one can study the Hawking radiation effect for various matter fields, and potentially implications on the ER=EPR. In the end we would like to elaborate more on the particle dynamics, and the average null energy condition (ANEC) using different test fields.

The paper is organized as follows. In Section \ref{Sec:Einstein-Rosen inspired BTZ wormhole}, we start from the BTZ black hole spacetime. We apply coordinate transformations to obtain an Einstein-Rosen BTZ wormhole spacetime. In Section \ref{Sec:The Einstein-Rosen BTZ wormhole requires exotic matter}, we show the presence of exotic matter at the wormhole throat. In Section \ref{Sec:Hawking radiation}, we show the effect of Hawking radiation in BTZ-like coordinates for scalar and vector fields, and in Section \ref{Sec:Hawking radiation-Kruskal} the Hawking radiation using Kruskal-like coordinates. In Section \ref{Sec:Possible implications on ER=EPR}, we point out the possible implications of ER=EPR. Further, in Section \ref{Sec:Particle dynamics}, we study the particle dynamics in the spacetime of BTZ wormhole. In Sections \ref{Sec:Average Null Energy Condition With Quantum Field}, we study the ANEC for scalar and vector test fields in BTZ coordinates and we point out the apparent divergence at the wormhole throat. In Sections \ref{ANEC IN KRUSKAL-LIKE}, we resolve this problems by using Kruskal-like coordinates which are non-singular at the wormhole horizon.  
Finally, in Section \ref{Sec:Conclusions}, we comment on our findings. 
\section{Einstein-Rosen inspired BTZ wormhole}\label{Sec:Einstein-Rosen inspired BTZ wormhole}

\quad Start with the BTZ black hole spacetime metric \cite{Banados:1992wn}, which is derived from the Einstein action 
\begin{equation}
    S=\int  \sqrt{-g} \left( \frac{R}{2 \kappa }-2\Lambda \right) d^3x \ ,
\end{equation}
where the cosmological constant $\Lambda=-1/\ell^2$ in our notation. From the above action, we can find the Einstein field equations 
\begin{equation}
    G_{\mu \nu}+\Lambda \, g_{\mu \nu}=\kappa \, T_{\mu \nu} \ .
\end{equation} 
The line element of the black hole in cylindrical coordinates reads
\begin{equation}\label{BTZ Metric}
    ds^2=-\left( \frac{r^2}{\ell^2}-M \right)dt^2+\frac{dr^2}{\frac{r^2}{\ell^2}-M}+r^2 d\phi^2 \ .
\end{equation}
Here $M$ is the ADM mass. This spacetime has a horizon at $r_h=\ell \sqrt{M}$. Let us perform a coordinate change 
\begin{equation}
    u^2=r-\ell \sqrt{M} \ ,
\end{equation}
we obtain a new spacetime metric given by 
\begin{equation}\label{Metric in u-cordinate}
    ds^2 = -\frac{u^2 (u^2 + 2 u_0)}{\ell^2} \, dt^2 + \frac{4 \ell^2 \, du^2}{u^2 + 2 u_0} + (u^2 + u_0)^2 \, d\phi^2 \ .
\end{equation}
where $u_{0}=\ell \sqrt{M}$ is the size of the wormhole throat.
One may notice in this coordinate system that $u$ 
\begin{equation}
    u=\pm \sqrt{r-\ell \sqrt{M}} \ ,
\end{equation}
will be real value for $r> \ell \sqrt{M}$ and will be imaginary for $r< \ell \sqrt{M}$. As $u$ varies from $-\infty $ to $\infty $, one finds $r$ varies from $+\infty$ to $\ell \sqrt{M}$ and then from $\ell \sqrt{M}$ to $+\infty$. In that sense, the $3-$dimensional spacetime can be described by two congruent sheets that are connected by a hyperplane at $r=\ell \sqrt{M}$, and that hyperplane is the so-called ``bridge''. Thus, Einstein and Rosen interpreted mass as a bridge in the spacetime.
First, we point out and observe an interesting fact about the radial null curves in the wormhole metric \eqref{Metric in u-cordinate} by setting $ds^2=d\phi=0$, yielding
\be
\frac{du}{dt}=\pm \frac{u (u^2+ 2\ell \sqrt{M})}{2 \ell^2 } \ .
\ee
The above quantity defines the “coordinate speed of light” for the wormhole metric, and as we can see there is a horizon with a coordinate location $u_h=0$ yielding 
\be
\left.\frac{du}{dt}\right|_{u_h=0}\longrightarrow 0~.
\ee
The surface area at the horizon in coordinates $u_h$ is given by
\begin{equation}
    A=\int \sqrt{g} d\phi=2 \pi u_0=2 \pi \ell \sqrt{M} \ .
\end{equation}
Although the metric \eqref{Metric in u-cordinate} has some nice properties, however, there seems to appear a small issue with metric \eqref{Metric in u-cordinate} if we compute the determinant 
\begin{eqnarray}\label{Det in ucoordinate}
    \det ||g_{\mu \nu}||=4 u^2 (u+u_0)^2 \ ,
\end{eqnarray}
which goes to zero, i.e., $ \det ||g_{\mu \nu}||=0$ when $u=u_h=0$. In addition to this problem, at $u=0$ the wormhole metric \eqref{Metric in u-cordinate} does not satisfy the Einstein field equations. This issue for the original Einstein-Rosen metric was pointed out in  \cite{Guendelman:2009er}.  To see the argument in our paper, we can use the Levi-Civita identity
\begin{eqnarray}
    R^t_t=-\frac{1}{\sqrt{-g_{tt}}}\nabla^2_{(2)}\left(  \sqrt{-g_{tt}} \right) \ ,
\end{eqnarray}
where $\nabla^2_{(2)}$ is the 2-dimensional spatial Laplacian. The metric \eqref{Metric in u-cordinate} solves the Einstein field equations for all $u \neq 0$, however at $u=0$, since $\sqrt{-g_{tt}} = u \sqrt{u^2+2 u_0}/\ell \sim u$, from the  Levi-Civita identity we get 
\begin{eqnarray}
    R^t_t  \sim \delta (u^2) \ ,
\end{eqnarray}
where $\delta (u)$ is the Dirac delta function. This result shows that we need a string with an energy-momentum tensor given by the Dirac delta function located on the common horizon $u=0$ the wormhole “throat”.  Namely the Einstein field equations should read 
\begin{equation}
    G_{\mu \nu}+\Lambda g_{\mu \nu}=\kappa T^{\rm string}_{\mu \nu} \ .
\end{equation} 
where ${T^{\rm string}}^{\mu}_\nu \sim T \, \delta (u) \, \rm diag(-1, 0, 0)$ located at the wormhole throat.  In the present paper, we can solve the issue related to Eq. \eqref{Det in ucoordinate} by using the following transformation in metric \eqref{BTZ Metric}
\begin{eqnarray}
    r^2=|u|+u_0^2=|u|+\ell^2 M \ ,
\end{eqnarray}
where $u \in (-\infty, \infty)$. The resulting metric reads
\begin{equation} \label{thirdmetric}
    ds^2 = -\frac{|u|}{\ell^2} \, dt^2 + \frac{ \ell^2 \, du^2}{4 |u|(|u| + \ell^2 M)} + (|u| + \ell^2 M) \, d\phi^2 \ . 
\end{equation}
 To the best of our knowledge, this Einstein-Rosen BTZ metric is new and has not been previously introduced in the literature. This metric  describes two identical copies of BTZ black hole  which correspond to the exterior region $u>0$ and $u<0$ and which are “glued” together at the horizon $u=0$. In particular we will focus on the spacetime region $u\geq 0$, and again $u_{0}=\ell \sqrt{M}$ is the size of the wormhole throat. For the the determinant of the metric in this case we get
\begin{eqnarray}
    \det ||g_{\mu \nu}||=\frac{1}{4}>0 \ .
\end{eqnarray}
Furthermore, by checking the temporal component of the metric, we can see if the wormhole is traversable or not, meaning we should not have any region of spacetime where $g_{tt} \rightarrow 0$ or becomes negative, ensuring that there are no horizons. 
The temporal component of our metric is 
\begin{equation}
	g_{tt} = - \frac{|u|}{\ell^2} \ .
\end{equation}
Setting $g_{tt}=0$, we get $u=0$. To determine if $u=0$ represents a true horizon, we can check the radial component of our metric,
\begin{equation}
	g_{uu}|_{u \rightarrow 0}=\frac{\ell^2 }{4|u|(|u|+\ell^2 M)} \rightarrow \infty \ .
\end{equation}
Alternatively, we can see this fact from the radial null curves in the wormhole metric \eqref{thirdmetric} by setting $ds^2=d\phi=0$, yielding
\be
\frac{du}{dt}=\pm \frac{2 u \sqrt{u^2+ \ell^2 M}}{ \ell^2 }  \ .
\ee
which as we pointed out  defines the “coordinate speed of light” for the wormhole metric, and as we can see, there is a horizon with a coordinate location $u_h=0$ yielding 
\be
\left.\frac{du}{dt}\right|_{u_h=0}\longrightarrow 0~.
\ee
The surface area at the horizon in coordinates $u_h$ is given by
\begin{equation}
    A=\int \sqrt{g} d\phi=2 \pi \sqrt{u+ \ell^2 M}\large|_{u_h=0}=2 \pi \ell \sqrt{M} \ .
\end{equation}
To further classify the singularity we need to compute curvature invariant. Calculating Ricci scalar and Kretschmann Scalar, in the region $u \neq 0$, we get,
\begin{equation}
	R = - \frac{6}{l^2}, \quad \quad K= \frac{12}{l^4} \ .
\end{equation}
To clarify this further, one can show that the metric \eqref{thirdmetric} is smooth everywhere except at the horizon located at $u=0$. At the horizon, the metric is only continuous but not differentiable. To see this, let us now calculate $R^t_t$ by considering Levi-Civita identity \cite{Guendelman:2009er}, from our metric \eqref{thirdmetric} we have
\begin{equation}
	R_t^t = -\frac{1}{\sqrt{-g_{tt}}} \frac{1}{\sqrt{h}} \frac{\partial}{\partial u } \left(\sqrt{h} h^{uu} 
 \frac{\partial}{\partial u } \sqrt{-g_{tt}}\right) \ ,
\end{equation}
However, due to the fact that there will be a term $\sqrt{-g_{tt}} \to 0 $ as $u \to 0$, and further containing second order derivative terms like
\begin{equation}
    \frac{\partial^2 }{\partial u^2} \sqrt{|u|} \sim  \frac{\partial }{\partial u} \left[  \frac{1}{2 \sqrt{|u|}} \left( \frac{\partial |u|}{\partial u}   \right)\right] \sim \frac{\partial^2 }{\partial u^2} |u| \sim  2\delta (u) \ ,
\end{equation}
From the perspective of the Einstein field equations, this demonstrates that a matter source is needed at $u=0$. In simple terms, the presence of matter at $u=0$ will lead to a contribution coming from the r.h.s of the Einstein field equation. The most general form of the energy-momentum tensor given by
\begin{eqnarray}
    T^{\mu \nu}=S^{\mu \nu } \delta(u) \ ,
\end{eqnarray}
where ${S^\mu}_\nu=(-\rho, P, P)$, with an equation of state $P=\omega \rho$, one can write the Einstein field equation
\begin{eqnarray}
    R_{\mu \nu}=8 \pi \left(S_{\mu \nu}-\frac{1}{2}g_{\mu \nu} S^\alpha_\alpha\right) \delta(u) \ ,
\end{eqnarray}
where the trace $S^\alpha_\alpha=-\rho+2P$. We can calculate the $R^t_t$ component from Levi-Civita idenity
\begin{eqnarray}\notag
R^t_t &=& -\frac{2}{\ell^2 u} \Big((\ell^2 + 2u) \frac{d}{du}|u| - (\ell^2 + u) \left(\frac{d}{du}|u|\right)^2 \\
&+& u (\ell^2 + u) \frac{d^2}{du^2}|u|\Big)  .
\end{eqnarray}
If we suppose that $T^{\mu \nu} =S^{\mu \nu} \delta(u)$ and $S^\alpha_\alpha = -\rho +2P$, we can calculate 
\begin{align}
    S_{tt} &= g_{tt} S^t_t = \big( -\frac{\abs{u}}{\ell^2} \big) (-\rho) \nonumber \\
    S_{uu} &= g_{uu} S^u_u = \frac{\ell^2}{4 \abs{u}(\abs{u}+\ell^2M} P \nonumber \\
    S_{\phi \phi} &= g_{\phi \phi} S^\phi_\phi = \big( \abs{u} +\ell^2 M \big) P \ .
\end{align}
We can calculate $R_{\mu \nu}$ to be 
\begin{align}
    R_{tt} &= 8 \pi \left( \frac{\abs{u} \rho}{2 \ell^2} + \frac{\abs{u} P}{\ell^2} \right) \delta(u)\nonumber \\
    R_{uu} &= 8 \pi \left( \frac{2 \ell^2 P}{4 \abs{u} \left(\abs{u}+\ell^2M\right)} + \frac{\ell^2 \rho}{2 4 \abs{u} \left( \abs{u} +\ell^2 M \right)} \right) \delta(u)  \nonumber \\
    R_{\phi \phi} &= 8 \pi \left( \frac{1}{2} \rho \left( \abs{u} +\ell^2 M \right) \right) \delta(u) \ .
\end{align}

For the region, $u\neq 0$, the spacetime has no curvature singularity that could block traversal through the wormhole, and having Kretschmann Scalar finite, too, confirms that there are no hidden singularities in the geometry.  Let us calculate the string tension for our metric \eqref{thirdmetric}. Conical singularities typically appear when the radial coordinate approaches zero or some specific value. We can analyse the behavior of the metric near $u=0$. Near $u=0$, the angular part of \eqref{thirdmetric} becomes 
\begin{equation}
	(u+\ell^2 M ) d\phi^2 \approx \ell^2 M d\phi^2 \ .
\end{equation}
Calculating the effective circumference we obtain
\begin{equation}
	C= 2 \pi \ell \sqrt{ M} \ .
\end{equation}
The deflecting angle $\delta \phi$ measures how much of the circumference diviates from $2\pi$, which we calculate to be 
\begin{equation}
	\Delta \phi = 2 \pi - \frac{2 \pi}{\ell\sqrt{ M}} \ ,
\end{equation}
from which we can calculate the tension to be
\begin{equation}
	T= \frac{1}{4G} \left(\frac{\ell\sqrt{ M}-1}{\ell\sqrt{ M}} \right).
\end{equation}
This equation shows that the string's tension is related to the wormhole throat $u_0=\ell \sqrt{M}$. For a positive string tension, i.e., $T>0$, we need the condition $\ell\sqrt{ M}>1$. Otherwise, we get a negative value for the string tension, i.e., $T<0$, for $\ell\sqrt{ M}<1$.  

\section{The Einstein-Rosen BTZ wormhole requires exotic matter}\label{Sec:The Einstein-Rosen BTZ wormhole requires exotic matter}
In this section, we aim to further elaborate on the wormhole geometry and demonstrate the presence of exotic matter at the wormhole's throat. To simplify the calculations, it is convenient to express our metric in Kruskal-like coordinates, allowing us to derive the energy density of the string matter. Let us write our metric in terms of Tortoise coordinate $u_*$ by introducing 
\begin{equation}
	\frac{d u_*}{du} = \frac{\ell^2}{2 u \sqrt{u+\ell^2 M}} \ ,
\end{equation}
and defining Kruskal-like coordinates $U$ and $V$ in terms of Tortoise Coordinate $u_*$ and time $t$ as 
\begin{equation}\label{U_V definition}
	U = -e^{-\frac{(t-u_*)}{2 \ell}} \quad \quad V= e^{\frac{(t+u_*)}{2 \ell}} \ .
\end{equation}
In Kruskal-like coordinates, our metric takes the form  
\begin{equation} \label{Kruskal-like metric}
	ds^2= -\frac{4 \ell^2}{(UV+ \ell^2M)}dUdV +(UV+\ell^2 M) d \phi^2 \ ,
\end{equation}
for the matter string let us take the  Polyakov action for a string is given by
\begin{equation} \label{stringaction}
S^{\rm string} = -\frac{T}{2} \int d\tau d\sigma \sqrt{-h} \, h^{ab} \, g_{\mu\nu} \, \partial_a X^\mu \, \partial_b X^\nu \ ,
\end{equation}
where $T$ is the string tension, $\tau$ and $\sigma$ are the are the worldsheet coordinates. Let stake $\tau=U$ and $\sigma=V$ and fix $\phi=\phi_0$. We need to get the energy density, for this, the energy-momentum tensor if found from
\begin{eqnarray}
    T^{\rm string}_{\mu \nu}=-\frac{2}{\sqrt{-g}}\frac{\delta S^{\rm string}}{\delta g^{\mu \nu}} \ .
\end{eqnarray}
Given the Kruskal-like metric \eqref{Kruskal-like metric}
we aim to find the energy density \( \rho^{\text{string}} \) using the energy-momentum tensor of the string.	
The energy-momentum tensor for the string derived from the Polyakov action is:	
\begin{equation}
T_{\mu \nu}^{\text{string}} = T \int d\tau d\sigma \left[ -(\partial_\tau X_\mu)(\partial_\tau X_\nu) + (\partial_\sigma X_\mu)(\partial_\sigma X_\nu) \right] \ .
\end{equation}
For the specific Kruskal-like coordinates \( (U, V) \), we focus on the component \( T_{UV} \):	
\begin{equation}
T_{UV} = T \int d\tau d\sigma \, (\partial_\tau X_U)(\partial_\tau X_V) 
\, .
\end{equation}	
The energy density \( \rho^{\text{string}} \) in Kruskal-like coordinates is given by:
\begin{equation}
\rho^{\text{string}} = g_{UV} T^{UV} \ .
\end{equation}	
Since the metric component is:	
\begin{equation}
g_{UV} = -\frac{4 \ell^2}{UV + \ell^2 M},
\end{equation}	
we can substitute this into the expression for the energy density:	
\begin{align} \label{rhostring}
\rho^{\text{string}} &= -\frac{4 \ell^2}{UV + \ell^2 M} T \int d\tau d\sigma \, (\partial_\tau X_U)(\partial_\tau X_V), \\
&= -\frac{4 T \ell^2}{UV + \ell^2 M} \int d\tau d\sigma \, (\partial_\tau X_U)(\partial_\tau X_V) \ .
\end{align}
 As $u \to 0$ the $UV \to 0$ \eqref{rhostring} now becomes 
\begin{equation}
    \rho^{\text{string}} \approx -\frac{4 T}{ M} \int d\tau d\sigma (\partial_\tau X_U)(\partial_\tau X_V) \ .
\end{equation}
Setting $ \int d\tau \sigma(\partial_\tau X_U)(\partial_\tau X_V)=1$, we get 
\begin{equation} \label{limit_rhostring}
    \rho^{\text{string}} \approx -\frac{4 T}{M} \ .
\end{equation}
The string negligible density at the wormhole throat suggest minimal influence on wormhole stability, passing through it without disturbing the spacetime geometry.
What one can do is to include Gaussian regularization to smooth out the behavior at the throat. In addition note that setting the equation of state parameter to zero, i.e., $\omega=0$ we get for the energy momentum-tensor as was expected ${T^{\rm string}}^{\mu}_\nu \sim \rho^{\text{string}}  \, \delta (u) \, \rm diag(-1, 0, 0)$. The linearized Einstein's equation in Lorentzian gauge is 
\begin{equation} \label{linearized}
\nabla^2 h_{\mu \nu}= -16 \pi T_{\mu \nu} \ .
\end{equation}
Using \eqref{limit_rhostring} we can write \eqref{linearized} as 
\begin{equation} \label{linearized1}
    \frac{1}{u}  \frac{d}{d u }\left( u \frac{d h_{00}}{du} \right)  = \frac{64 \pi T}{M \epsilon \sqrt{\pi}} e^{-\frac{u^2}{\epsilon^2}}, 
\end{equation}
where we use the Gaussian regularization given as 
\begin{equation}
    \delta(u) = \frac{1}{\epsilon \sqrt{\pi}} e^{-\frac{u^2}{\epsilon^2}}.
\end{equation}
Then for our $T_{\mu \nu}$ we can write 
\begin{equation}
    T^{\textrm{string} 0}_0  \sim \frac{4T}{M} \delta(u)
\end{equation}
and $T_0^0= g_{00} T^{ \text{string} 0 }_0 $. For our \eqref{linearized} we get
\begin{equation}
\frac{1}{u} \frac{d}{du} \left( u \frac{d h_{00}}{du} \right) = -\frac{u}{\ell^2} \frac{4T}{M} \frac{1}{\epsilon \sqrt{\pi}} e^{-\frac{-u^2}{\epsilon^2}}
\end{equation}
which can be solved to obtain 
\begin{equation}
    h_{00}(u) = \left( C_1 - \frac{T \epsilon^3 \sqrt{\pi}}{M \ell^2} \right) \ln{u} +C_2
\end{equation}
and by requiring that $h_{00} \to 0$ for $u \to \infty$ we can set $C_2=0$ and $C_1 - \frac{T \epsilon^3 \sqrt{\pi}}{M \ell^2}= 0$, where for the solution now we can write 
\begin{equation}
    h_{00}(u) =\frac{T \epsilon^3 \sqrt{\pi}}{M \ell^2} \ln{\frac{u_0}{u}}
\end{equation}
where $u_0$ is some reference point introduced implicitly such that $\ln{u_0}=0$.\newline
\begin{figure}[ht]
    \centering
    \includegraphics[width=0.85\linewidth]{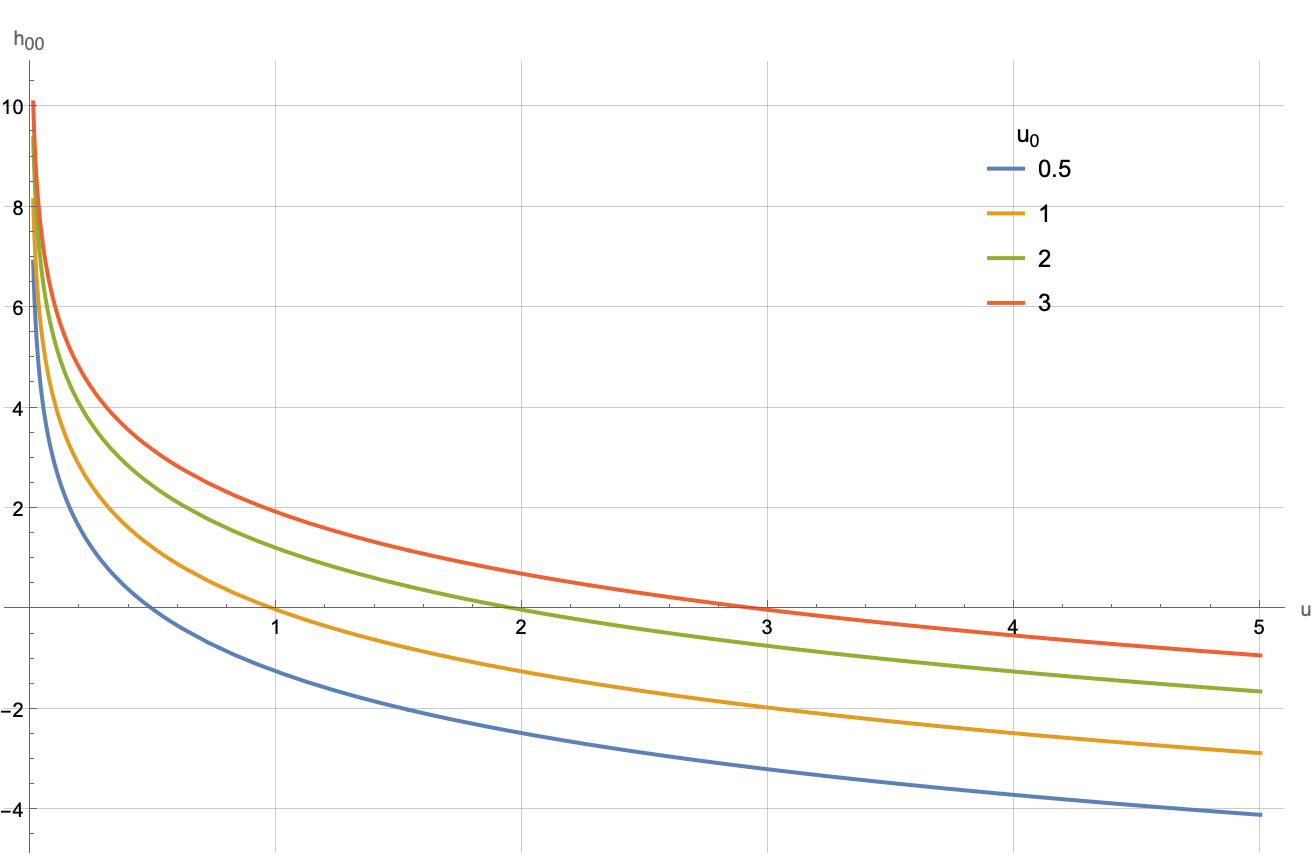}
    \caption{Gravitational potential as a function of  $u$.}
    \label{potential}
\end{figure}

What we can see is that by introducing exotic matter we can have transeversable wormhole but with a strong gravitational potential where $u_0$ sets a scale for the gravitational potential which defines a region where the transversable path of a wormhole is less affected by the strong gravitational force near the throat. As $u>u_0$, the logarithmic term becomes negative, suggesting that the region beyond $u_0$ exhibits an attractive gravitational field, as the potential becomes more negative as you move away from the source, atypical behavior in gravitational systems. This fact can be seen from the plot of the gravitational potential given in Fig. \eqref{potential}.
On the other hand for $u <u_0$ we are in a region  dominated by the effects of exotic matter, keeping the wormhole open where the exotic matter is providing necessary repulsive force to keep the wormhole open.

\section{Hawking radiation in BTZ-like coordinates}\label{Sec:Hawking radiation}
The presence of the horizon implies that the quantum tunneling of particles from ``another universe'' to our universe can form Hawking radiation and, consequently, detecting particles by a distant observer located in our universe. We can study the tunneling of different massless or massive spin particles; and in the present work, we focus on studying the tunneling of scalar $\Phi$ and vector particles $\Psi^{\mu}$. 
\subsection{Scalar field}
First let us consider the case of tunneling of the scalar particle from the wormhole spacetime. We start with the Klein-Gordon equation [with $c=1$]
\begin{equation}
\frac{1}{\sqrt{-g}}\partial_{\mu}\left( \sqrt{-g}g^{\mu\nu}\partial_{\nu} \Phi\right) -\frac{m^{2}}{\hbar^{2}}\Phi=0 \ ,  
\end{equation}
then using the wormhole metric \eqref{thirdmetric} we get
\begin{eqnarray}
    &-&\frac{\ell^2}{u} \frac{\partial^2 \Phi}{\partial t^2}+\frac{1}{M \ell^2+u}\frac{\partial^2 \Phi}{\partial \phi^2}\\\notag
    &+&\frac{4 u (M \ell^2+u)}{\ell^2}\frac{\partial^2 \Phi}{\partial u^2}+\frac{4 (M \ell^2+2u)}{\ell^2}\frac{\partial \Phi}{\partial u}-\frac{m^2 \Phi}{\hbar^2}=0 \ .
\end{eqnarray}
By making use of the WKB ansatz (see for example \cite{Shankaranarayanan:2000qv})
\begin{equation}
\Phi=\exp\left(\frac{i}{\hbar}S(t,u,\phi)\right) \ ,
\end{equation}
where, in general, one can assume the form of action $S(t,u,\phi)$ in a powers of $\hbar$ as follows
\begin{equation}
    S(t,u,\phi)=S_{0}(t,u,\phi)+\sum_{i=1} \hbar^i S_{i}(t,u,\phi).
\end{equation}
Taking into the consideration the symmetries of the metric \eqref{thirdmetric} given by three corresponding Killing vectors $(\partial/\partial_{t})^{\mu}$ and
$(\partial/\partial_{\phi})^{\mu}$, the action as the following form
\begin{equation}
S_0(t,u,\phi)=-E t+R(u)+j \phi,
\end{equation}
where $E$ is the energy of the particle, and $j$ denotes the angular momentum of the particle corresponding to the angles $\phi$. 
As noted in \cite{Parikh:1999mf}, the WKB approximation can be justified due to the following argument: It is expected that the typical wavelength of the radiation to be of the order of the size of the horizon, however, when the outgoing wave is traced back towards the horizon, its wavelength as measured by an static observers is increasingly blue-shifted. Near the horizon, the radial wavenumber approaches infinity and the point particle, or WKB, approximation is justified. From the above equations in the leading order terms in $\hbar$,  we obtain the following equation 
\begin{eqnarray}\notag
    &4& u^2 (M \ell^2+u)^2 \left( \frac{\rm d}{{\rm d}u} R(u)\right)^2+\ell^2 \Big[-m^2 u^2+E^2 M \ell^4\\
    &+& u \left(\ell^2 (E^2-M m^2)-j^2\right)\Big]=0 \ ,
\end{eqnarray}
from where we get the radial part $R(u)$ as follows
\begin{equation}
R_{\pm}=\pm \int \frac{ \ell \sqrt{(E^2\ell^2-m^2 u)(u+\ell^2 M)-j^2u}} {2 u (u+\ell^2 M)}du \ .
\end{equation}
To solve the above integral, let us first introduce the function
\be \label{F(u)}
\mathcal{F}(u)=\frac{2 u \sqrt{u+\ell^2 M}}{\ell^2}=\mathcal{F}'(u)|_{u=0} (u-u_h)+\cdots \ .
\ee
and rewrite the above solution as 
\begin{equation}
R_{\pm}=\pm \int \frac{ \sqrt{E^2-u\left[\frac{m^2}{\ell^2}+\frac{j^2}{\ell^2 (u+\ell^2 M)}\right]}} {\mathcal{F}(u)}du \ .
\end{equation}
Now, there is a singularity in the above integral when $u_h=0$, meaning that $\mathcal{F}\to 0$. So in order to find the Hawking temperature, we now make use of the equation
\be
\lim_{\epsilon \to 0} \text{Im}\frac{1}{u-u_h\pm i \epsilon }=\delta(u-u_h)~,
\ee
where $u_h=0$. In this way we find 
\be
\text{Im}R_{\pm}=\pm \frac{ E \pi }{\mathcal{F}'(u)|_{u=0}}~.
\ee
Using $p_u^{\pm}=\pm \partial_u R_{\pm}$, for the total tunneling rate gives
\begin{align}\notag
\Gamma &=\exp\left(\frac{1}{\hbar}\text{Im} (E \Delta t^{\rm out,in})-\frac{1}{\hbar}\text{Im} \oint p_{u} \mathrm{d}u\right)\\
&=\exp\left(-\frac{4 E \pi }{\mathcal{F}'(u)|_{u=0}}\right) \ .
\end{align}
where we shall again fix $\hbar =1$ we have also added a temporal part contribution due to the connection of the interior region and the exterior region of the wormhole via $t \to t - i \pi /\mathcal{F}'(u)$. We can finally obtain the Hawking temperature for the wormhole by using the 
Boltzmann factor $\Gamma=\exp(-E/T_H)$, and setting $\hbar$ to unity, so that it results with 
\be\label{Hawking_temperature_S}
    T_H=\frac{\mathcal{F}'(u)|_{u_h=0}}{4 \pi }=\frac{\sqrt{M}}{2 \pi \ell}~.
\ee

\subsection{Vector field}
One can also study the motion of a massive vector particle of mass $m$, described by the vector field with the corresponding action 
\be
    S=-\int d^4x\sqrt{g}\left( \frac{1}{2}\Psi_{\mu\nu}\Psi^{\mu\nu}+\frac{m^2}{\hbar^2}\Psi_{\mu}\Psi^{\mu}\right)~.
\ee
The Proca equation (PE), which reads 
\be\label{Proca}
\nabla_{\mu}\nabla^{[\mu}\Psi^{\nu]}-\frac{m^{2}}{\hbar^{2}}\Psi^{\nu}=\frac{1}{\sqrt{-g}}\partial_{\mu}\left[\sqrt{-g}\partial^{[\mu}\Psi^{\nu]}\right]-\frac{m^{2}}{\hbar^{2}}\Psi^{\nu}=0 \ ,
\ee
with
\be
\nabla_{[\mu}\Psi_{\nu]}=\frac{1}{2}(\nabla_{\mu}\Psi_{\nu}-\nabla_{\nu}\Psi_{\mu}):=\Psi_{\mu\nu}~.
\ee

Solving tunneling equations exactly is quite hard. So, we apply the WKB approximation method 
\be\label{WKB}
\Psi_{\nu}=A_{\nu}\exp\left[\frac{i}{\hbar}\left(S_{0}(t,u,\phi)+\sum_i \hbar^i S_{i}(t,u,\phi)\right)\right]~,
\ee
and we can choose the ansatz for the action given by Eq. (IV.5). If we keep only the leading order of $\hbar$, we find a set of four differential equations. These equations can help us to construct a $3\times 3$ matrix $\aleph$, which satisfies the following matrix equation 
\be
 \aleph(A_{1},A_{2},A_{3})^{T}=0~.
\ee
with the matrix elements
\begin{eqnarray}\notag
    \aleph_{11}&=& \aleph_{23}=-4 E (M \ell^2+u) \left(\frac{\rm d}{{\rm d} u}R(u)\right)\ ,\\\notag
     \aleph_{21} &=& \frac{-4E^2M \ell^4-4u(-Mm^2+E^2)\ell^2+4m^2u^2+4j^2u}{\ell^2}\ ,\\\notag
      \aleph_{31} &=& \aleph_{22}=-\frac{4 u j }{\ell^2}\left(\frac{\rm d}{{\rm d} u}R(u)\right),\\\notag
      \aleph_{12} &=& \aleph_{33} = -\frac{E j \ell^2}{u (M \ell^2+u)}\ ,\\\notag
      \aleph_{32} &=&  \frac{4 u^2 (M \ell^2+u) \left(\frac{\rm d}{{\rm d} u}R(u)\right)^2+ \ell^2(m^2 u-\ell^2 E^2)}{\ell^2 (M \ell^2+u) u} \ ,\\\notag
      \aleph_{13} &=& \frac{-4 u (M \ell^2+u)^2 \left(\frac{\rm d}{{\rm d} u}R(u)\right)^2-\ell^2 m^2 (M \ell^2+u)-\ell^2 j^2}{u (M \ell^2+u)}\ .
\end{eqnarray}
From the matrix relation, we get the equation 
\begin{eqnarray}\notag
    &4& u^2 (M \ell^2+u)^2 \left( \frac{\rm d}{{\rm d}u} R(u)\right)^2+\ell^2 \Big[-m^2 u^2+E^2 M \ell^4 \\
    &+&u \left(\ell^2 (E^2-M m^2)-j^2\right)\Big]=0 \ ,
\end{eqnarray}
We solve for the radial part to get the following integral 
\be
R_{\pm}=\pm \int \frac{ \sqrt{E^2-u\left[\frac{m^2}{\ell^2}+\frac{j^2}{\ell^2 (u+\ell^2 M)}\right]}} {\mathcal{F}(u)}du~,
\label{R as function of u}
\ee
and $\mathcal{F}(u)$ is given by Eq. (IV.8). We see that the radial solution is the same as in the case of scalar field, hence going through  the same calculations we get the same result for the Hawking temperature
given by Eq. (IV.23).

\section{Hawking radiation in Kruskal-like coordinates}\label{Sec:Hawking radiation-Kruskal}
\subsection{Scalar field}
We saw in Section III, the Kruskal-like coordinates are important to regularize the apparent divergence which can arise in BTZ spacetime wormhole near the horizon. In this section we would like to show that indeed one can compute also the Hawking radiation via tunneling using the Kruskal-like coordinates. This shows the invariance of the Hawking temperature under coordinate systems. We start from the metric Eq.\eqref{Kruskal-like metric} with the ansatz's
\begin{equation}\label{Scalar Field Phi}
    \Phi(U,V,\phi)=\exp \left[\frac{i}{\hbar} \left(S_{0}(U,V,\phi)+\sum_i \hbar^i S_{i}(U,V,\phi) \right)   \right]
\end{equation}
where 
\begin{eqnarray}
    S_{0}(U,V,\phi)=R(U,V)+j \phi.
\end{eqnarray}
Using WKB approximation from the Klein-Gordon equation we get
\begin{equation}
 (\partial_U R)(\partial_V R) (Ml^2+UV)^2-l^2 \left( m^2 (Ml^2+UV)+j^2   \right)=0.
\end{equation}
To find the particles energy we need the Killing
vector $\xi^\mu$ for the Kruskal-like coordinates. One can check that one such vector field  
\begin{eqnarray}
    \xi^\mu=\left(-U/\ell,V/\ell, 0\right)
\end{eqnarray}
is a solution to the Killing equation $ \nabla_{(\mu} \xi_{\nu)}=0$. Therefore, for the energy we have the following relation
\begin{eqnarray}
    \tilde{E}=E \ell=-\xi^\mu \partial_\mu S_0=U \partial_U R-V \partial_V R.
\end{eqnarray}

From the last two equations we solve for $\partial_U R$ and obtain two solutions
\begin{equation}
  \partial_U R_{\pm}=\frac{\tilde{E} \pm \sqrt{\tilde{E}^2-4 UV l^2 \left( \frac{m^2}{Ml^2+UV}- \frac{j^2}{(Ml^2+UV)^2}   \right)}}{2U}.
\end{equation}
At the horizon $UV \to 0$, we see that there are two solutions 
\begin{equation}
  \lim_{UV \to 0}  \partial_U R_-=0,\,\,\,\,\,  \text{and}\,\,\,\,\,   \lim_{UV \to 0} \partial_U R_+=\frac{\tilde{E}}{U}.
\end{equation}

The zero contribution is explained by the fact the the metric is regular  at the horizon the ingoing particle/light experiences no barrier and hence no contribution to the tunneling effect, however, the outgoing particle/light experiences barrier and we can assign e tunneling probability. In a similar way we have 
\begin{equation}
  \partial_V R_{\pm}=\frac{-\tilde{E} \pm \sqrt{\tilde{E}^2-4 UV l^2 \left( \frac{m^2}{Ml^2+UV}- \frac{j^2}{(Ml^2+UV)^2}   \right)}}{2V},
\end{equation}
at the horizon $UV \to 0$, we also have two solutions 
\begin{equation}
  \lim_{UV \to 0}  \partial_V R_+=0,\,\,\,\,\,  \text{and}\,\,\,\,\,   \lim_{UV \to 0} \partial_V R_-=-\frac{\tilde{E}}{U}.
\end{equation}
In order to find the temperature let us see first the imaginary  contribution to the action given by
\begin{equation}
    S_0=R+j \phi=\int \partial_U R_+ dU +\int \partial_V R_+ dV + j \phi
\end{equation}
Only the first term in the action has a pole at the horizon, while the second and third term will have a zero contribution to the imaginary  part. We can write
\begin{eqnarray}
    \Im R= \Im R_+=\Im \int \frac{\tilde{E}}{UV}d\left(UV\right).
\end{eqnarray}
To solve this integral, we have multiplied and divided by $V$ (the coordinate $V$ near the horizon behaves as constant).  Then using the new variable $Z=UV$ which, can be further written in terms of $u_\star$ as $Z=-\exp(u_\star/\ell)$, gives
\begin{equation}
 \Im R_+= \Im \int \frac{ \tilde{E} \,du_\star}{\ell}=\Im \int \frac{ E \ell^2 du}{2 u \sqrt{u+\ell^2 M}}.
\end{equation}
We can easily solve the above integral, in fact we can define $\mathcal{F}(u)$ as in Eq. (IV.8) and gives the same contribution to the imaginary part, i.e., $\text{Im}R_+= E \pi /\mathcal{F}'(u)|_{u=0}$. In general we have to consider the total loop contribution, hence we also have to find the imaginary  contribution to the action given by
\begin{equation}
    S_0=\int \partial_U R_- dU +\int \partial_V R_- dV + j \phi.
\end{equation}
In this case, only the second term has a contribution to the imaginary  part. Hence
\begin{eqnarray}
    \Im R= \Im R_-=-\Im \int \frac{\tilde{E}}{UV}d\left(UV\right).
\end{eqnarray}
Note that in this case in solving the above integral we have multiplied and divided by $U$ (this time the coordinate $U$ near the horizon behaves as constant). Or, alternatively, in terms of the coordinate $u$, we can write
\begin{equation}
 \Im R_-= -\Im R_+=-\frac{E \pi}{\mathcal{F}'(u)|_{u=0}}.
\end{equation}

For the total loop contribution we have 
\begin{equation}
    \text{Im} \oint p_{u} \mathrm{d}u= 2\Im R_+=\frac{2E \pi}{\mathcal{F}'(u)|_{u=0}},
\end{equation}
This leads to the total tunneling rate given by
\begin{eqnarray}\notag
    \Gamma &=&\exp \left(\frac{1}{\hbar}\text{Im} (E \Delta t^{\rm out,in}) -\frac{2}{\hbar} \Im R_+\right)\\
    &=&\exp\left(-\frac{4 E \pi }{\mathcal{F}'(u)|_{u=0}}  \right),
\end{eqnarray}
where we set again $\hbar=1$ and we have included also the temporal contribution to the tunneling rate (as was explained bellow Eq. (IV.12)). Finally, we can use the Boltzmann factor $\Gamma=\exp(-E/T_H)$ and we get the same expression for the Hawking temperature given by Eq. (IV.23).

\subsection{Vector Field}
Let us turn now our attention to solve the Proca equation that describes the massive vector fields by using 
\be\label{WKB}
\Psi_{\nu}(U,V,\phi)=A_{\nu}\exp\left[\frac{i}{\hbar}\left(S_0+\sum_i \hbar^i S_{i}(U,V,\phi)\right)\right]~.
\ee
where, the action given by Eq. (V.2). If we keep only the leading order of $\hbar$, we can construct a $3\times 3$ matrix $\aleph$, which satisfies a matrix equation $ \aleph(A_{1},A_{2},A_{3})^{T}=0$, with the matrix elements given by
\begin{eqnarray} \notag
    \aleph_{11}&=&-\frac{(M \ell^2+UV)^2}{4 \ell^4} \left(\partial_V R\right)^2,\\ \notag
     \aleph_{21} &=&  \aleph_{12} =\frac{2 (M \ell^2+UV)^2 \left(\partial_V R\right) \left(\partial_U R\right)}{8 \ell^2}\\\notag
     &-& \frac{4 \ell^2 \left(m^2 (M \ell^2+UV) +j^2 \right)}{8 \ell^2} ,\\ \notag
      \aleph_{31} &=& \aleph_{13}=\frac{j }{2\ell^2}\left(\partial_V R\right),\\ \notag
      \aleph_{22} &=& -\frac{ (M \ell^2+UV)^2}{4 \ell^4} \left(\partial_U R\right)^2,\\ \notag
      \aleph_{32} &=&  \aleph_{23}=  \frac{j }{2\ell^2}\left(\partial_U R\right).
\end{eqnarray}
From the above matrix relation, we get the equation 
\begin{equation}
 (\partial_U R)(\partial_V R) (Ml^2+UV)^2-l^2 \left( m^2 (Ml^2+UV)+j^2   \right)=0.
\end{equation}
If we now use this relation and the equation for the energy of the particle/light we get
\begin{equation}
\partial_V R= -\frac{\tilde{E}}{V}+\frac{U}{V}\partial_U R.
\end{equation}
By solving this equation we again obtain two solutions for the radial part 
\begin{equation}
  \partial_U R_{\pm}=\frac{ \tilde{E} \pm \sqrt{ \tilde{E}^2-4 UV l^2 \left( \frac{m^2}{Ml^2+UV}- \frac{j^2}{(Ml^2+UV)^2}   \right)}}{2U}.
\end{equation}
At the horizon, $UV \to 0$, the only non-zero contribution comes from the outgoing waves (see Eq. (V.7)). This situation is exactly the same as in the case of scalar particles: the non-zero contribution arises from the outgoing particle. We can omit the details, as they lead to the same result for the imaginary part, given by Eq. (V.10). Thus, we arrive at the same expression for the Hawking temperature. This demonstrates once again the invariance of Hawking radiation with respect to the choice of coordinate system and the spin of the fields.

\section{Possible implications on ER=EPR}\label{Sec:Possible implications on ER=EPR}

We follow an analysis where we consider another example of entangled particles that could lead to wormhole formation: a pair of particles with extreme mass, capable of undergoing gravitational collapse into a black hole. In our case, this could manifest as two maximally entangled BTZ black holes, potentially representing an Einstein-Rosen BTZ wormhole. Such a spacetime has horizon (at the wormhole throat) $u_0=\ell \sqrt{M}$. The entropy can be found by means of the Hawking-Bekenstein relation $S=A/4= \pi u_0/2$. It is interesting to see that this wormhole temperature coincides with the Hawking temperature of the BTZ black hole given by the metric \eqref{BTZ Metric}. In other words, the role of the black hole horizon is played by the wormhole throat. As an example, we can consider a pair created in a maximally entangled state connected by an ER bridge with the spin state (see for example \cite{Dai:2020ffw,Jusufi:2023dix}) 
\begin{equation}
\ket {\Psi} = \frac{1}{\sqrt{2}}\left(\ket{\uparrow_1 \downarrow_2}+\ket{\downarrow_1 \uparrow_2}\right),
\end{equation}

These particle-black holes are entangled. From the first law of black hole thermodynamics, we can find the Bekenstein-Hawking entropy related to our wormhole metric, or the BTZ black hole metric \eqref{BTZ Metric}. The entropy at the horizon reads
\begin{eqnarray}
    S=\frac{\pi \ell \sqrt{M}}{2}.
\end{eqnarray}

On the other hand, the entropy must be larger than
the entanglement entropy between a pair given by
\begin{eqnarray}
    S=-k_B \Tr \left[ \hat{\rho} \ln(\hat{\rho}) \right],
\end{eqnarray}
where $\hat{\rho}$ is the reduced density operator. For our case, using the last equation it is easy to show that
\begin{eqnarray}\label{Entropy_2}
    S= N\ln(2),
\end{eqnarray}
where we have set $k_B=1$. This leads to the expression for the wormhole throat 
\begin{eqnarray}
    u_0=\ell \sqrt{M}=\frac{2 N \ln(2)}{\pi}.
\end{eqnarray}

These simple calculations show that, using Bekenstein-Hawking entropy and entanglement entropy, we obtain a value for the size of the wormhole throat that is proportional to the number of quantum bits, $N$,  multiplied by $\ln(2)$.

\section{Particle dynamics }\label{Sec:Particle dynamics}
In the following section we consider the motion of  a test particle of mass $m_0$ in a gravitational background given by the reparametrization-invariant world-line action
\begin{equation}
	S =\frac{1}{2} \int d\lambda \left[ \frac{1}{e} g_{\mu \nu} \dot{x}^\mu \dot{x}^\nu - e m_0^2  \right],
\end{equation}
where $\dot{x}^\mu = \frac{dx^\mu}{d \lambda}$ where $\lambda$ is the world line reparametrization and $e$ is world-line "einbein" and in our case $ x^\mu =(t,u,\theta)$. Writing down the Lagrangian we have 
\begin{equation} \label{Lagrangian}
	L = \frac{1}{2e} \Bigg( -\frac{u}{\ell^2} \dot{t}^2 + \frac{\ell^2}{4 u (u + \ell^2 M)} \dot{u}^2 + (u + \ell^2 M) \dot{\phi}^2 \Bigg) - \frac{1}{2} e m_0^2
\end{equation}
where we use the metric \eqref{thirdmetric}
Due to time translational invarianvce and rotational symmetry as a conserve quantities we have Energy $E$ and angular momentum $L$, repsecively. 
\begin{equation}
	\dot{t}=\frac{E \ell^2 e}{u},
\end{equation}
and 
\begin{equation}
	\dot{\phi} = \frac{L e}{u +\ell^2 M}.
\end{equation}
For our radial equation using the mass-shell condition we have,
\begin{equation} \label{radialeq}
	\dot{u}^2 + V_{\text{eff}}(u) = E^2,
\end{equation}
where the effective potential is given by 
\begin{equation}
	V_{\text{eff}}(u) = \frac{4uL^2 e^2}{\ell^2} + \frac{4u(u+\ell^2 M)m_0^2}{\ell^2} -4E^2 e^2 (u+\ell^2 M )
\end{equation}
If we have a test particle, using  we can write  
\begin{equation} \label{integralforu1}
	t(u) = t_0 + \int_{u_0}^{u} \frac{E \ell^2 e\, du}{u \sqrt{4 E^2 e^2 (u + \ell^2 M) - \frac{4 u L^2 e^2}{\ell^2} - \frac{4 u (u + \ell^2 M) m_0^2}{\ell^2}}},
\end{equation}

\begin{equation}
	\tau(u) = \tau_0 + \int_{u_0}^{u} \frac{du}{\sqrt{4 E^2 e^2 (u + \ell^2 M) - \frac{4 u L^2 e^2}{\ell^2} - \frac{4 u (u + \ell^2 M) m_0^2}{\ell^2}}}.
\end{equation}
For large values of $u$, we can simplify \eqref{integralforu1}, from which we show that 
\begin{equation}
	t(u)-t_0 \approx - \ell^2 \left( \frac{1}{\sqrt{u}} - \frac{1}{\sqrt{u_0}}  \right)
\end{equation}
What we can see is that at large $u$, far from horizon, the laboratory time $t(u)$ grows slowly, depending on the value of $u_0$ which suggests that the time seen by an external observer approaches a constant value.\newline
For small values of $u$ (near the throat), the behavior depends heavily on the structure of the potential and the angular momentum term might act as a barrier to create a repulsive potential. The integral \eqref{integralforu1} simplifies to 
\begin{equation}
	t(u)-t_0 \approx \frac{\ell^2}{ 2 \sqrt{M}}  \ln(u)
\end{equation}
As we can see, for small values of $u$ the time $t$ goes to infinity and it appears that, as seen from an external observer, it will take an infinite amount of time for a particle to reach the horizon.\newline
To understand the behavior of the particle, we need to relate the laboratory time \( t \) (the time experienced by an observer in the lab) to the proper time \( \tau \) (the time experienced by the particle). The integral relationship between \( t \) and \( \tau \) is given by:
\begin{equation}
	\frac{dt}{d\tau} = \frac{E \ell^2 e}{u(\tau)}.
\end{equation}
To proceed, we analyze the behavior of \( u(\tau) \) near the throat and far from it.
Near the throat of the wormhole, the radial coordinate \( u \) approaches zero. In this regime, we approximate the expression for \( u(\tau) \) as:
\begin{equation}
	u(\tau) = 2 E e \ell \sqrt{M} (\tau - \tau_0) + u_0,
\end{equation}
where \( \tau_0 \) is the initial proper time and \( u_0 \) is the initial value of \( u \) at \( \tau = \tau_0 \). Substituting this into the relation for \( \frac{dt}{d\tau} \), we obtain:
\begin{equation}
	\frac{dt}{d\tau} = \frac{\ell}{2 \sqrt{M} (\tau - \tau_0)}.
\end{equation}
This shows that \( \frac{dt}{d\tau} \) diverges as \( \tau \to \tau_0 \), meaning that the coordinate time \( t \) grows without bound as the particle approaches the throat.
The integral for \( t(\tau) \) near the throat gives:
\begin{equation}
	t(\tau) = t_0 + \frac{\ell}{\sqrt{M}} \ln( (\tau - \tau_0)+ \frac{u_0}{2Ee\ell \sqrt{M}}).
\end{equation}
Thus, the coordinate time \( t \) diverges logarithmically as the particle approaches the throat, implying that from the laboratory frame, the particle appears to take an infinite amount of time to reach the throat.\newline
Far from the throat, as \( u \to \infty \), we approximate the behavior of \( u(\tau) \) as:
\begin{equation}
	u(\tau) \sim \tau^2.
\end{equation}
Substituting this into the expression for \( \frac{dt}{d\tau} \), we find:
\begin{equation}
	\frac{dt}{d\tau} \sim \frac{E \ell^2 e}{\tau^2}.
\end{equation}
Integrating this, we obtain:
\begin{equation}
	t(\tau) \approx t_0 + \frac{E \ell^2 e}{\tau_0} - \frac{E \ell^2 e}{\tau}.
\end{equation}
As \( \tau \to \infty \), the term \( \frac{E \ell^2 e}{\tau} \) vanishes, and the laboratory time \( t(\tau) \) approaches a constant value. This indicates that far from the throat, the particle's time evolution slows down in the laboratory frame.\newline
For an observer in the laboratory frame, the behavior of the test particle near the wormhole throat is intriguing. As the particle approaches the throat, the observer sees the particle’s motion slow down progressively, appearing to take an infinite amount of time $t$ to reach the throat. This effect resembles the phenomenon observed near the event horizon of a black hole, where a distant observer perceives the particle as 'frozen' near the horizon, though here it is the wormhole's throat that causes the delay in the particle's approach.

However, from the perspective of the test particle, proper time \( \tau \) continues to evolve normally. The particle experiences no dramatic slowdown and crosses the throat smoothly in a finite amount of proper time. This is a key distinction between a traversable wormhole and a black hole: in the case of the wormhole, the particle can pass through the throat without encountering an event horizon or singularity.\newline
We can also study the violation of null energy condition (NEC). Let us suppose that we have a null vector with a general form given as  $k^\mu = (k^t, k^u, k^\phi)$. In our case we assume that we have no angular momentum component, setting $k^\phi=0$. Since the null vector must satisfy the condition $g_{\mu \nu} k^\mu k^\nu =0$, we can write
\begin{equation} \label{geodesic for vectors}
	\frac{u}{\ell^2} (k^t)^2 = \frac{\ell^2 }{4u(u+\ell^2M)}(k^u)^2.
\end{equation}
Solving in terms of $k^t$, we have:
\begin{equation}
	k^t= \frac{\ell^2}{ 2u\sqrt{u+\ell^2 M}} k^u.
\end{equation}
Now we can write our null vector as 
\begin{equation} \label{nullvector}
	k^\mu = \left(\frac{\ell^2}{2u\sqrt{u+\ell^2 M}} k^u, k^u, 0 \right).
\end{equation}
The null energy condition requires $T_{\mu \nu} k^\mu k^\nu >0$, or in terms of components and using null vector \eqref{nullvector}, we can write 
\begin{equation}
	T_{\mu \nu} k^\mu k^\nu = T_{tt} (k^t)^2 + T_{uu} (k^u)^2 \geq 0 
\end{equation}
which we calculate to be 
\begin{equation} \label{energymomentumnullradial}
T_{\mu \nu} k^\mu k^\nu = 0
\end{equation}
 
Note that there is a second possibility if we rewrite Eq.\eqref{geodesic for vectors} 
\begin{equation}
k^u=\frac{2 u \sqrt{u+\ell^2 M}}{\ell^2} k^t,
\end{equation}
yielding
\begin{equation} \label{nullvector2}
	k^\mu = \left(k^t, \frac{2 u \sqrt{u+\ell^2 M}}{\ell^2} k^t, 0 \right),
 \end{equation}
 resulting with
 \begin{equation} \label{energymomentumnulltime}
T_{\mu \nu} k^\mu k^\nu = 0
\end{equation}
We can see from \eqref{energymomentumnullradial} and \eqref{energymomentumnulltime} that NEC is satisfied for null vectors \eqref{nullvector} and \eqref{nullvector2}. 
which solution is also very expected because in this region we have vacuum solution.\newline
One can also check the same calculation for massive particles satisfying $u^\mu u_\mu=-1$. Starting from our metric \eqref{thirdmetric}, for purely radial motion we have
\begin{equation} \label{massive particle condition}
	-\frac{u}{\ell^2} (u^t)^2 + \frac{\ell^2}{4u(u+\ell^2 M)} (u^u)^2= -1
\end{equation}
from which we can derive 
\begin{equation}
	u^t = \frac{\ell}{\sqrt{u}} \sqrt{1+\frac{\ell^2}{4u(u+\ell^2 M)} (u^u)^2},
\end{equation}
and now for our first vector we have
\begin{equation} \label{vector1}
	u^\mu=\left(u^t,u^u,u^\phi\right) = \left(  \frac{\ell}{\sqrt{u}} \sqrt{1+\frac{\ell^2 (u^u)^2}{4u(u+\ell^2 M)} }, u^u, 0   \right).
\end{equation}
From \eqref{massive particle condition} we can also derive
\begin{equation} \label{vector2}
	u^u = \pm \sqrt{ \frac{4u(u+\ell^2 M)}{\ell^2} \left(  \frac{u}{\ell^2} (u^t)^2 -1   \right) },
\end{equation}
For \eqref{vector1} we calculate that 
\begin{equation} \label{NECmassive1}
	T_{\mu \nu} u^\mu u^\nu = 0,
\end{equation}
while for \eqref{vector2} we calculate that 
\begin{equation} \label{NECmassive2}
T_{\mu \nu} u^\mu u^\nu  = 0.
\end{equation}


\section{Average Null Energy Condition with Quantum Field}\label{Sec:Average Null Energy Condition With Quantum Field}

\subsection{Scalar Field}
In this subsection, we study the Average Null Energy Condition of a free scalar field of mass $m$, denoted as $\Phi(x)$, the corresponding action is expressed as
\begin{eqnarray}\notag
	    S &=& \int d^3x \sqrt{-g}\Big(-\frac{1}{2}g^{\mu\nu}\partial_\mu \Phi(x) \partial_\nu \Phi(x) -\frac{1}{2} m^2 \Phi^2(x) \\
     &-&\frac{1}{4} \lambda \Phi^4(x) \Big) \ ,
	\end{eqnarray}
here we usually take $\lambda > 0$. The stress-energy tensor is obtained by varying the action with respect to the metric $g^{\mu\nu}$ as
\begin{equation}\label{Stress-Energy Scalar}
    T_{\mu\nu} = \nabla_\mu \Phi \nabla_\nu \Phi - \frac{1}{2} g_{\mu\nu}  \left( \left[g^{\delta \rho } \nabla_\delta \Phi  \nabla_\rho \Phi \right] + m^2\Phi^2 +\frac{\lambda}{2}\Phi^2 \right) \ .
\end{equation}
Here we choose the $\Phi$ as a function of $t,\,u$ and $\phi$ and the simplest choice we can start with is 
\begin{equation}\label{General Phi}
    \Phi(t,u,\phi) = \exp\left[\frac{i}{\hbar} \left(S_{0}(t,u,\phi)+\sum_i \hbar^i S_{i}(t,u,\phi) \right) \right]
\end{equation}
Now, using \eqref{General Phi} we can compute the expression in square parenthesis of \eqref{Stress-Energy Scalar} as
\begin{eqnarray}\label{Square Parenthesis}\notag
   g^{\rho\sigma} \partial_{\rho} \Phi \partial_{\sigma} \Phi &=& \Big\{ \frac{E^2 \ell^2}{u \hbar^2} - \frac{4 u \,(u+\ell^2 M)}{\ell^2 \hbar^2}\left(\frac{dR(u)}{du}\right)^2 \\
   &-&  \frac{j^2}{\hbar^2\,(u+\ell^2 M)} \Big\} \Phi^2  \ . 
\end{eqnarray}
As already discussed in \eqref{R as function of u}, the form of $\frac{dR(u)}{du}$ is
\begin{equation}\label{dR/du}
   \frac{dR(u)}{du}=  \frac{ \ell \sqrt{(E^2\ell^2-m^2 u)(u+\ell^2 M)-j^2u}} {2 u (u+\ell^2 M)}
\end{equation}

However do to the presence of the coordinate singularity at the horizon, we can write
\begin{equation}
  \lim_{u\to 0} \frac{dR(u)}{du} = \frac{E}{2 \ell \sqrt{M}}\delta (u),
 \end{equation}
the first term is regular in the region $u \neq 0$, while the second term has an apparent divergence due to the coordinate singularity at the horizon $u=0$. 
Using \eqref{Stress-Energy Scalar}, \eqref{Square Parenthesis} and \eqref{dR/du} we have
\begin{eqnarray}\notag
 {T}_{t\,t}  &=& \Big[ - \frac{E^2 }{2 \hbar^2} - \frac{2 u^2 \, (u+\ell^2 M)}{\ell^4 \hbar^2}\left(\frac{dR(u)}{du}\right)^2 \\\notag
 &-& \frac{j^2 u}{2\ell^2\, (u+\ell^2 M)} + \frac{m^2 u}{2\ell^2} + \frac{\lambda u}{4\ell^2} \Phi^2 \Big] \Phi^2 \ .   \nonumber \\\notag
{T}_{u\,u} &=& \Bigg[ - \frac{E^2 \ell^4 }{8 u^2\, (u+\ell^2 M) \hbar^2} - \frac{1}{2 \hbar^2}\left(\frac{dR(u)}{du}\right)^2 \\\notag
&+& \frac{j^2 \ell^2}{8 u (u+\ell^2 M)^2 \hbar^2}- \frac{m^2 \ell^2}{8 u\, (u+\ell^2 M)} \\\notag
&-& \frac{\lambda \ell^2}{16 u \, (u+\ell^2 M)}\Phi^2 \Bigg] \Phi^2  \ . \nonumber \\ 
 {T}_{t\, u} &=& \left[  \frac{E }{\hbar^2} \left(\frac{dR(U)}{du} \right)\right] \Phi^2  \ . \nonumber
\end{eqnarray}

\begin{figure}[ht]
	\begin{center}
		\includegraphics[scale=0.60]{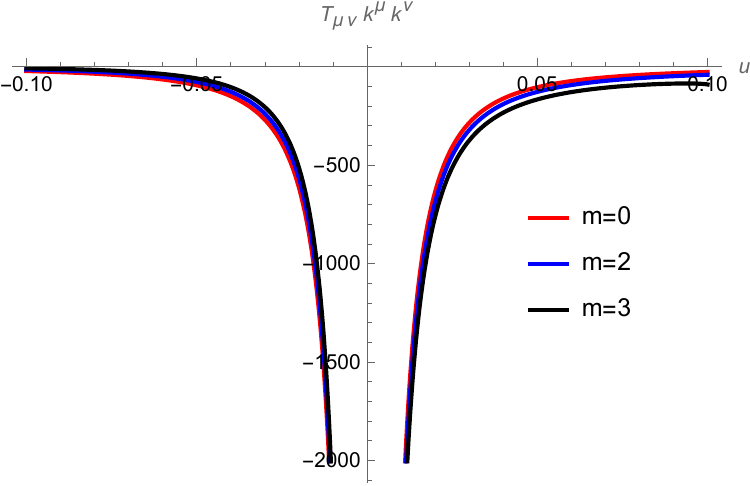}
  \hspace{0.1Cm}
  \includegraphics[scale=0.60]{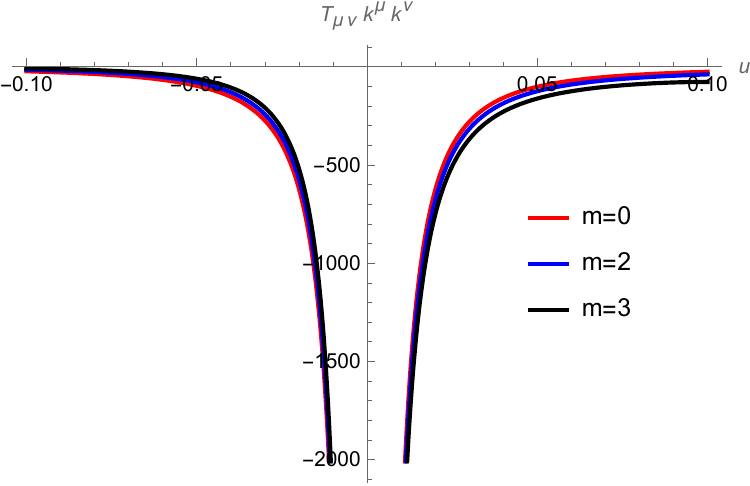}
	\end{center}
	\caption{Plot of $k^\mu k^\nu T_{\mu \nu}$ vs $u$}
	\label{fig:Double null component of scalar field}
\end{figure}

By choosing the null vector as 
\begin{equation*}
    k^\mu = \left(k^t,\, \frac{2 u \sqrt{u+\ell^2 M}}{\ell^2}k^t, \, 0\right) \ .
\end{equation*}
From the Lagrangian in Eq.\eqref{Lagrangian}, we can easily find the form of $k^t$   
\begin{eqnarray}
    k^t=\dot t\equiv\frac{dt}{d\lambda}= \frac{ {-p_t}}{u/\ell^2}=\frac{E \ell^2}{u} \ ,
\end{eqnarray}
where we have used $p_t=-E$. Now, using this we can have the form of null vector as 
\begin{equation}\label{Kmu general}
k^\mu = \left(\frac{E \ell^2}{u} ,\, 2 E \sqrt{u+\ell^2 M}, \, 0\right)     \ .
\end{equation}
Using the expression of $k^\mu$ and stress-energy tensor, the expression for double null-component is 
\begin{eqnarray}\notag
    k^\mu k^\nu T_{\mu \nu} &=& \frac{1}{4 \hbar^2 u^2 \left(\ell^2 M+u\right)^{3/2}} \Bigg[ E^2 \ell^2 \Bigg(4 E \ell \left(\ell^2 M+u\right) \\\notag
    &&\sqrt{\left(\ell^2 M+u\right) \left(E^2 \ell^2-m^2 u\right)-j^2 u} -5 E^2 \ell^2 \\\notag
    &&\left(\ell^2 M+u\right)^{3/2}+u \sqrt{\ell^2 M+u}\\
    & \times & \left(j^2+m^2 \left(\ell^2 M+u\right)\right)\Bigg)\Bigg] \Phi^2 \ .
\end{eqnarray}
We see that the energy density exhibits a pronounced sensitivity, in fact it should be viewed as an apparent singularity due to the singular nature of coordinates.  This divergence can be seen in the limit $u\rightarrow 0$, i.e., at the throat of the wormhole. We can verify the same by plotting them as shown in Fig. \eqref{fig:Double null component of scalar field}. 


\subsection{Vector Field}

\begin{figure}[ht]
	\begin{center}
		\includegraphics[scale=0.60]{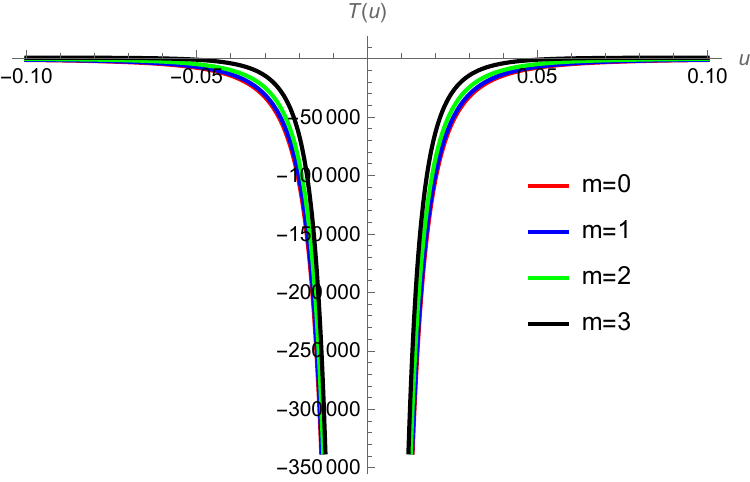}
  \hspace{0.1Cm}
  \includegraphics[scale=0.60]{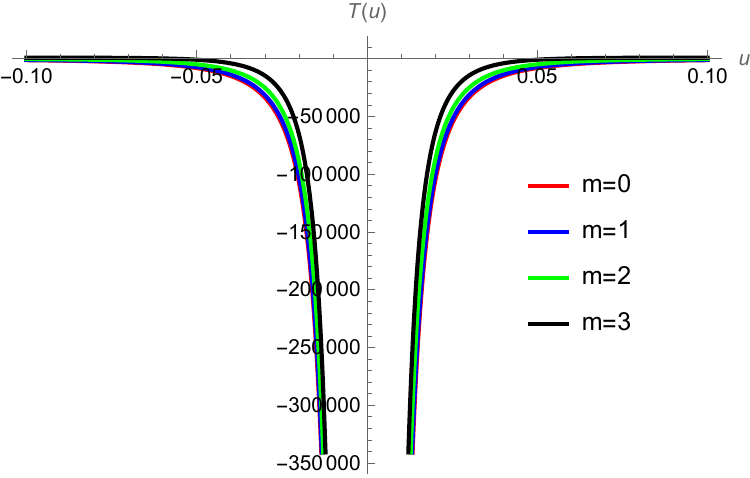}
	\end{center}
	\caption{Plot of $k^\mu k^\nu T_{\mu \nu}$ vs $u$ for massive vector field}
	\label{fig:Double null component of vector field}
\end{figure}

This subsection will do similar analyses with the massive vector field. The action for a massive vector field is 
\begin{equation} \label{Massive action}
    \mathcal{S}= \int d^4x \sqrt{-g} \left(-\frac{1}{4} g^{\alpha \beta} g^{\tau \nu} F_{\alpha \tau}F_{\beta \nu} -\frac{1}{2} m^2  g^{\tau \nu} \Psi_\tau \Psi_\nu \right) \ ,
\end{equation}
where $F_{\tau \nu}=\nabla_\tau \Psi_\nu - \nabla_\nu \Psi_\tau$. The corresponding stress energy is 
\begin{eqnarray}
     T_{\mu\nu} &=& g^{\alpha \beta} F_{\alpha \mu}F_{\beta \nu}+m^2 \Psi_\mu \Psi_\nu \\\notag
     &-&\frac{1}{2} g_{\mu\nu}  \left[\frac{1}{4}g^{\gamma \delta}g^{\sigma \rho} F_{\gamma \sigma} F_{ \delta \rho}+\frac{1}{2}m^2g^{\gamma \delta} \Psi_\gamma \Psi_\delta \right]   \ .
\end{eqnarray}
here, similarly, we can use
\begin{equation}\label{Vector field Psi}
    \Psi_\mu(t, u, \phi) =  A_\mu \exp\left[\frac{i}{\hbar} \left(S_{0}(t,u,\phi)+\sum_i \hbar^i S_{i}(t,u,\phi) \right) \right]
\end{equation}
Since we aim to compute the double null component of the stress-energy tensor, we can directly ignore the term with $g_{\mu \nu}$. So we have 
\begin{eqnarray}\label{Vector field stress energy}
     T_{\mu\nu} = g^{\alpha \beta} F_{\alpha \mu}F_{\beta \nu}+m^2 \Psi_\mu \Psi_\nu  \ .
\end{eqnarray}
The components of \( F_{\mu\nu} \) of the interest (define $\mathcal{E}=e^{\frac{i}{\hbar} \left( -Et + R(u) + j\phi \right)}$
\begin{eqnarray}
    F_{ut} &=& \frac{i \mathcal{E}}{\hbar} \left( E A_1 + \frac{\partial R(u)}{\partial u} A_0 \right),\\
    F_{\phi t} &=& \frac{-i \mathcal{E}}{\hbar} (E A_2 + j A_0),\\
    F_{\phi u} &=&  \frac{i \mathcal{E}}{\hbar}  \left(   j A_1 - A_2 \frac{\partial R(u)}{\partial u}\right) \ . 
\end{eqnarray}
Now, with the help of this again, we can find the component of stress-energy tensor (Here $\frac{\partial R(u)}{\partial u}=R'(u)$)
\begin{eqnarray}\notag
    T_{tt} &=& -\frac{\mathcal{E}^2}{\hbar^2} \Big[\frac{4 u (u+\ell^2 M) \left( E A_1 + R'(u) A_0 \right)^2}{\ell^2 } \\\notag
    &+&\frac{(E A_2 + j A_0) ^2}{u+\ell^2 M} - m^2 A_0^2 \hbar^2 \Big] \\\notag
     T_{uu} &=& -\frac{\mathcal{E}^2}{\hbar^2} \Big[-\frac{\ell^2  \left( E A_1 + R'(u) A_0 \right)^2}{u }  + \frac{(j A_1 - R'(u) A_2) ^2}{u+\ell^2 M} \\\notag
     &-& m^2 A_1^2 \hbar^2 \Big] \\\notag
     T_{ut} &=& -\frac{\mathcal{E}^2}{\hbar^2} \Big[ \frac{1}{u+\ell^2 M} (j A_1 - R'(u) A_2) (E A_2 + j A_0) \\\notag
     &-&m^2 \hbar^2 A_0 A_1 \Big] \ . 
\end{eqnarray}
With the help of the choice of the null vector in Eq.\eqref{Kmu general}, the double-null component of the stress-energy tensor is
\begin{eqnarray}\nonumber
    k^\mu k^\nu T_{\mu \nu} &=& \frac{1}{4 u^3 (M+u)^2}\Bigg[-4 (M+u) (-A_0^2 m^2 u (M+u)\\\nonumber
    &+&\Big(A_0 \sqrt{\Xi}+2 A_1 u (M+u)\Big)^2+u (A_0+A_2)^2)\\\notag
    &+&4 u \sqrt{M+u} \Big(2 A_0 A_1 m^2 u (M+u)^2 -(A_0+A_2)\\\notag
    &\times& (2 A_1 u (M+u)-A_2 \sqrt{\Xi})\Big)+(M+u)\\\notag
    &\times& \left(A_0 \sqrt{\Xi}+2 A_1 u (M+u)\right)^2+4 A_1^2 m^2 u^3\\\notag
    & \times& (M+u)^3- u \Big(A_2 \sqrt{\Xi}- 2 A_1 u (M+u)\Big)^2\Bigg],
\end{eqnarray}
where $\Xi=\left(1-m^2 u\right) (M+u)-u$. The energy density shows a significant sensitivity, which manifests as an apparent singularity arising from the singular nature of the coordinates. This divergence becomes evident in the limit \( u \rightarrow 0 \), corresponding to the throat of the wormhole. We can confirm this behavior by plotting the result for $k^\mu k^\nu T_{\mu \nu}$  presented in Fig. \eqref{fig:Double null component of vector field}. As we will elaborate this apparent divergence is a result of the choice of the BTZ coordinates for our wormhole geometry used in our calculations. 

\section{Average Null Energy Condition in Kruskal-like coordinates}\label{ANEC IN KRUSKAL-LIKE}
In this final section we would like to solve the problem of apparent divergence that appeared in computing the double null-component computed in BTZ coordinates.  In particular, we shall calculate the stress-energy tensor in Kruskal-like coordinates with the metric in Eq.\eqref{Kruskal-like metric}. The null-vector has the following form
\begin{equation}
k^\mu=\left(k^U,k^V,0\right)
=\left(0,-\frac{E(UV+l^2M)}{4 l V},0\right),
\end{equation}
or
\begin{equation}
k^\mu=\left(k^U,k^V,0\right)
=\left(\frac{E(UV+l^2M)}{4 l U},0,0\right).
\end{equation}
In addition, we will use the derivative of $R_\pm(U)$, which satisfies the following conditions:
\begin{equation}
  \partial_U R_{\pm}=\frac{ \tilde{E} \pm \sqrt{ \tilde{E}^2-4 UV l^2 \left( \frac{m^2}{Ml^2+UV}- \frac{j^2}{(Ml^2+UV)^2}   \right)}}{2U} \ ,
\end{equation}
and
\begin{equation}
  \partial_V R_{\pm}=\frac{ -\tilde{E} \pm \sqrt{ \tilde{E}^2-4 UV l^2 \left( \frac{m^2}{Ml^2+UV}- \frac{j^2}{(Ml^2+UV)^2}   \right)}}{2V} \ .
\end{equation}

In what follows we are going to check the double null components using the above relation the scalar and vector fields. 

\subsection{Scalar field}
For the scalar field, the double-null component of the stress-energy tensor for the ingoing waves, it is easy to see that 
\begin{eqnarray}
    k^\mu k^\nu T_{\mu \nu} &=& (k^U)^2 T_{UU} + (k^V)^2 T_{VV} \ , \\\notag
   &=& -\frac{1}{\hbar^2}\left[\left(k^U\right)^2 \left( \partial_U R_-\right)^2 + (k^V)^2 \left( \partial_V R_+\right)^2 \right] \Phi^2
\end{eqnarray}
using the form of the scalar field given in Eq.\eqref{Scalar Field Phi}, along with the condition at the horizon $UV \to 0$,  for ingoing  waves, we have
\begin{equation}
\lim_{UV \to 0} \left\{(k^U)^2 T_{UU}+(k^V)^2 T_{VV}\right\} \to 0. 
\end{equation}
At the horizon $UV \to 0$, we have considered the result for the ingoing waves; in particular we need the choice $\partial_U R_-=0$ and $\partial_V R_+=0$, respectively. These results show that the final result is indeed finite and non-singular, as we expected. 

One might ask what happens in the case of outgoing waves, specifically when we consider $\partial_U R_+ = E / U$, along with the expression $\partial_V R_- = -E / V$ (with $U$ and $V$ going to zero at the horizon). It is not difficult to see that an apparent divergence arises. However, this result can be explained by the infinite redshift effect experienced by outgoing waves as observed from a distant observer.

\subsection{Vector field}
We turn our attention now to the vector field, namely we would like to compute the double null component of the stress-energy tensor given by 
\begin{eqnarray}
    k^\mu k^\nu T_{\mu \nu} &=& (k^U)^2 T_{UU} + (k^V)^2 T_{VV}  \ .
\end{eqnarray}
The expression for $T_{UU}$ and $T_{VV}$ can be calculated using the Eq.\eqref{Vector field stress energy}. Now, the expression of field strength for the calculation of $T_{UU}$ or $T_{VV}$ can be easily written by using Eq.\eqref{Vector field Psi}. The field strength expression of interests are
\begin{eqnarray}
    F_{\phi U} &=&  \mathcal{E}\frac{i}{\hbar}\left[A_U j - A_\phi \partial_U R(U) \right] \nonumber \\
    F_{\phi V} &=& \mathcal{E}\frac{i}{\hbar}\left[A_V j - A_\phi \partial_V R(U) \right] \ . \nonumber
\end{eqnarray}

Using them, we end up with the expression 
\begin{eqnarray}
    (k^U)^2 T_{UU} &=& -\frac{\Tilde{E}^2 \mathcal{E}^2(\ell^2M+UV)^2}{16 \ell^2 V^2} \Bigg[\frac{(A_U j-A_\phi \partial_U R)^2}{UV+\ell^2 M}\nonumber \\,
    && -A_U^2 m^2\hbar^2\Bigg] \\
     (k^V)^2 T_{VV} &=& -\frac{\Tilde{E}^2 \mathcal{E}^2(\ell^2M+UV)^2}{16 \ell^2 U^2} \Bigg[\frac{(A_V j-A_\phi \partial_V R)^2}{UV+\ell^2 M}\nonumber \\
    && -A_V^2 m^2\hbar^2\Bigg]
\end{eqnarray}

Let us explore some of the implications that can be deduced from the above relations. First, consider the case at the wormhole horizon, i.e., when  $UV \to 0$, and by considering the result for the ingoing waves: $\partial_U R_-=0$ and $\partial_V R_+=0$, we have 
\begin{eqnarray}
    k^\mu k^\nu  T_{\mu \nu} &=& -\frac{\Tilde{E}^2 \mathcal{E}^2 \ell^2M}{16 \ell^2 } \Bigg[\left\{\frac{A_U^2}{V^2}-\frac{A_V^2}{U^2}\right\} \left\{j^2-m^2 \hbar^2\right\} \Bigg]\ . \nonumber 
\end{eqnarray}

It is important to mention that in the case where $\partial_U R_-=0$, the coordinate $V$ near the horizon near the horizon remains constant. Similarly, when $\partial_V R_+=0$, the coordinate $U$ near the horizon also behaves as constant. This implies that the final result is finite and does not diverge, as expected. Furthermore, if we impose the condition $j=m=0$, the double null component vanishes, i.e., $k^\mu k^\nu T_{\mu \nu} \to 0$. Another possibility is of course to set $A_U/V=A_V/U$, which also results in a vanishing double null component.

In Kruskal-like coordinates, the ingoing fields does not experience the wormhole horizon, however the outgoing fields indeed experience the gravitational barrier due to the presence of the horizon. As in the case of scalar fields, for outgoing waves, $U$ and $V$ are zero at the horizon. Thus, for outgoing waves, for vector fields a divergence also appears, which can be explained by the fact that signals from such waves near the horizon undergo infinite redshift as observed by a distant observer.\\

\section{Conclusions}\label{Sec:Conclusions}
In this paper, we used the famous BTZ black hole solution and obtained a  novel metric, which we refer to as the Einstein-Rosen BTZ wormhole metric. We showed that the metric is a solution to the Einstein field equations with a negative cosmological constant and, importantly, it has a horizon at the throat, implying that our wormhole metric describes a one-way traversable wormhole. 

Due to the presence of the horizon, we were able to show the presence of with Hawking radiation as observed by an observer located at some distance from the wormhole. At the level of semiclassical approximations, we employed the tunneling picture and showed that the Hawking temperature is unchanged by the spin of the fields. It is also demonstrated that at the wormhole throat, the spacetime is not a solution to the Einstein field equations, but rather it contains an exotic string matter source with negative tension. It is interesting to mention that the energy density of the string is shown to depend on the string tension and wormhole mass and, in addition, the exotic matter is argued to stabilize the wormhole geometry. 

We also elaborated a possible implications on ER=EPR, in particular it is pointed out that the by means if Bekenstein-Hawking entropy in one hand and the entanglement entropy on the other hand, the size of the wormhole throat is proportional to the number of quantum bits, $N$, multiplied by $\ln(2)$.   

We elaborated in great details the particle dynamics. We found a similar phenomenon to that observed in the black hole case: as the observer approaches the wormhole, time appears to freeze from the perspective of an outside observer. However, the proper time measured by the approaching observer remains finite. This is analogous to what happens when a particle approaches the black hole horizon. In the final part we used the WKB approximations for scalar and vector test fields to test the ANEC. It is found that the quantity $k^\mu k^\nu T_{\mu \nu} $ in BTZ coordinates can diverge at the wormhole throat. We resolve this issue by using the Kruskal-like coordinates, in particular we show that $k^\mu k^\nu T_{\mu \nu} $ for ingoing scalar/vector waves is a finite quantity and tends to zero at the wormhole throat. For for outgoing waves, on the other hand, an apparent divergence appears, which is explained from the fact that outgoing waves near the horizon undergo infinite redshift as observed by a distant observer. Finally, the above picture holds within the WKB approximation method; however, it remains an open question whether exact solutions for test fields exist near the wormhole throat. We plan to investigate this issue further in the near future.

\end{document}